\documentclass[prd,nofootinbib,superscriptaddress,showpacs,preprint]{revtex4-1}
\usepackage[T1]{fontenc}
\usepackage{amstext,amssymb}
\usepackage{amsmath}
\usepackage{epsfig}
\graphicspath{{Figs/}}
\usepackage{graphicx}
\usepackage[usenames,dvipsnames]{color}
\usepackage{xspace}
\usepackage{braket}
\usepackage{units}
\usepackage{slashed}
\usepackage{color}

\newcommand{\del}{\partial}

\def \epsilon {\varepsilon} 

\newcommand{\hc}{\ensuremath{\text{h.c.}}}

\begin{document}
\title{Subdominant Left-Right Scalar Dark Matter as Origin of the 750 GeV Di-photon Excess at LHC}
\author{Debasish \surname{Borah}}
\email{dborah@iitg.ernet.in}
\affiliation{Department of Physics, Indian Institute of Technology Guwahati, Assam 781039, India}
\author{Sudhanwa \surname{Patra}}
\email{sudha.astro@gmail.com}
\affiliation{Center of Excellence in Theoretical and Mathematical Sciences,
Siksha 'O' Anusandhan University, Bhubaneswar-751030, India}
\author{Shibananda \surname{Sahoo}}
\email{shibananda@iitg.ernet.in}
\affiliation{Department of Physics, Indian Institute of Technology Guwahati, Assam 781039, India}

\begin{abstract}
We study the possibility of explaining the recently reported 750 GeV di-photon excess at LHC within the framework of a left-right symmetric model. The 750 GeV neutral scalar in the model is dominantly an admixture of neutral components of scalar bidoublets with a tiny fraction of neutral scalar triplet. Incorporating $SU(2)$ septuplet scalar pairs into the model, we enhance the partial decay width of the 750 GeV neutral scalar into di-photons through charged septuplet components in loop while keeping the neutral septuplet components as subdominant dark matter candidates. The model also predicts the decay width of the 750 GeV scalar to be around 36 GeV to be either confirmed or ruled out by future LHC data. The requirement of producing the correct di-photon signal automatically keeps the septuplet dark matter abundance subdominant in agreement with bounds from direct and indirect detection experiments. We then briefly discuss different possibilities to account for the remaining dark matter component of the Universe in terms of other particle candidates whose stability arise either 
due to remnant discrete symmetry after spontaneous breaking of $U(1)_{B-L}$ or due to high $SU(2)$-dimension forbidding their decay into lighter particles.
\end{abstract}
\maketitle

\section{Introduction}
After the discovery of the Standard Model (SM) Higgs boson in 2012, the Large Hadron Collider (LHC) experiment has been searching 
for new physics beyond the Standard Model (BSM). In spite of many theoretical motivations for BSM physics around the TeV scale, 
the LHC experiment has not been able to discover anything new so far. However, very recently the ATLAS and CMS collaborations  
\cite{atlasconf,lhcrun2a,CMS:2015dxe}  of the LHC have reported a di-photon resonance $pp \to X \to \gamma \gamma$ with an invariant mass of $750~$GeV at around $3\sigma$ level of confidence. The reported cross-sections for di-photon excess signal are given by
\begin{align} 
&\sigma_{\rm ATLAS}\left( pp \to X \right) \cdot {\rm Br} \left( X\to \gamma \gamma \right) \simeq (10\pm 3) \mbox{fb}\, , \nonumber \\
&\sigma_{\rm CMS}\left( pp \to X \right) \cdot {\rm Br} \left( X\to \gamma \gamma \right) \simeq (6\pm 3) \mbox{fb}\,. \nonumber
\end{align}
ATLAS collaboration has also hinted towards a large decay width of this 750 GeV scalar, around 45 GeV, though the CMS collaboration 
still prefers a narrow width. Although it is equally probable that the statistical significance of the di-photon signal may go down with more LHC data, it is nevertheless tantalising to consider the possibility that the excess signal is an indication of some interesting BSM physics. The Landau-Yang theorem forbids the possibility of 
spin-1 resonance as an interpretation of the di-photon signal \cite{Landau:1948,Yang:1950rg}. Interestingly, this excess signal can be easily interpreted by 
postulating a new spin-0 scalar particle with mass around $750~$GeV which can be produced at LHC and subsequently decay 
into two photons. 
Many attempts have already been made in this context 
in order to explain the di-photon excess events with the simple extension of SM singlet scalar along with new physics heavy scalars and vector-like fermions 
charged under SM gauge group and others including supersymmetric variants
\cite{DiChiara:2015vdm,Pilaftsis:2015ycr,Knapen:2015dap,Backovic:2015fnp,Molinaro:2015cwg,Gupta:2015zzs,
Ellis:2015oso,Higaki:2015jag,Mambrini:2015wyu,Buttazzo:2015txu,Franceschini:2015kwy,Angelescu:2015uiz,
Bellazzini:2015nxw,McDermott:2015sck,Low:2015qep,Petersson:2015mkr,
Cao:2015pto,Kobakhidze:2015ldh,Agrawal:2015dbf,Chao:2015ttq,Fichet:2015vvy,Curtin:2015jcv,Csaki:2015vek,
Aloni:2015mxa,Demidov:2015zqn,No:2015bsn,Bai:2015nbs,Matsuzaki:2015che,Dutta:2015wqh,Becirevic:2015fmu,
Cox:2015ckc,Martinez:2015kmn,Bian:2015kjt,Chakrabortty:2015hff,Ahmed:2015uqt,
Deppisch:2016scs,Hernandez:2016rbi,Dutta:2016jqn,Modak:2016ung,
Danielsson:2016nyy,Chao:2016mtn,Csaki:2016raa,Karozas:2016hcp,
Ghorbani:2016jdq,Han:2016bus,Ko:2016lai,Nomura:2016fzs,
Ma:2015xmf,Palti:2016kew,Potter:2016psi,Jung:2015etr,
Marzola:2015xbh,Falkowski:2015swt,
Nakai:2015ptz,Harigaya:2015ezk,
Ellwanger:2015uaz,Karozas:00640,Csaki:00638,Chao:00633,Danielsson:00624,Ghorbani:00602,
Ko:00586,Han:00534,Nomura:00386,Palti:00285,Potter:00240,Palle:00618,Dasgupta:2015pbr,Bizot:2015qqo,
Goertz:2015nkp,Kim:2015xyn,Craig:2015lra,Cheung:2015cug,Allanach:2015ixl,Altmannshofer:2015xfo,Huang:2015rkj,
Belyaev:2015hgo,Liao:2015tow,Chang:2015sdy,Luo:2015yio,
Kaneta:2015qpf,Hernandez:2015hrt,Low:2015qho,Dong:2015dxw,Kanemura:2015vcb,Kanemura:2015bli,Kang:2015roj,Chiang:2015tqz,
Ibanez:2015uok,Huang:2015svl,Hamada:2015skp,Anchordoqui:2015jxc,Bi:2015lcf,Chao:2015nac,Cai:2015hzc,Cao:2015apa,
Tang:2015eko,Dev:2015vjd,Gao:2015igz,Cao:2015scs,Wang:2015omi,An:2015cgp,Son:2015vfl,Li:2015jwd,Salvio:2015jgu,
Park:2015ysf,Han:2015yjk,Hall:2015xds,Casas:2015blx,Zhang:2015uuo,Liu:2015yec,Das:2015enc,Davoudiasl:2015cuo,Cvetic:2015vit,
Chakraborty:2015gyj,Badziak:2015zez,Patel:2015ulo,Moretti:2015pbj,Gu:2015lxj,Cao:2015xjz,Pelaggi:2015knk,Dey:2015bur,
Hernandez:2015ywg,Murphy:2015kag,deBlas:2015hlv,Dev:2015isx,Boucenna:2015pav,Kulkarni:2015gzu,Chala:2015cev,
Bauer:2015boy,Cline:2015msi,Berthier:2015vbb,Kim:2015ksf,Bi:2015uqd,Heckman:2015kqk,Huang:2015evq,Cao:2015twy,Wang:2015kuj,
Antipin:2015kgh,Han:2015qqj,Ding:2015rxx,Chakraborty:2015jvs,Barducci:2015gtd,Cho:2015nxy,Feng:2015wil,Bardhan:2015hcr,
Han:2015dlp,Dhuria:2015ufo,Chang:2015bzc,Han:2015cty,Arun:2015ubr,Chao:2015nsm,Bernon:2015abk,Carpenter:2015ucu,Megias:2015ory,
Alves:2015jgx,Gabrielli:2015dhk,Kim:2015ron,Benbrik:2015fyz,8thJan-1,8thJan-2,8thJan-3,8thJan-4,8thJan-5, last750}.

On the other hand, the left-right symmetric models (LRSM) \cite{lrsm,lrsmpot} originally motivated for 
natural explanation for maximal parity violation in weak interactions as well as to provide 
light neutrino masses via seesaw mechanisms have the potential to explain 
the di-photon signal when extended with additional vector like fermions \cite{Dasgupta:2015pbr,Cao:2015xjz,Dev:2015vjd,Deppisch:2016scs}. 
The added advantage of considering left-right symmetric models with vector like fermions is that it can be easily embedded 
in a grand unified theory unifying all three known fundamental interactions. One interesting possibility is to interpret the $750$~GeV di-photon excess signal in the simple extensions of two Higgs doublet models where one of the lightest neutral Higgs scalars (apart from $125$~GeV SM Higgs boson) is considered as the $750$ GeV resonance produced sufficiently via gluon gluon fusion process and subsequently decaying to two photons via one-loop diagrams containing charged components of high $SU(2)$-dimension scalars in loop \cite{Han:2015qqj}. Thus, the di-photon signal cross-section $\sigma(pp \to H \to \gamma \gamma)$ 
can be increased so as to fit with the ATLAS and CMS data with the inclusion of additional scalar having high $SU(2)-$dimension. 

With these motivations, in the present work, we consider the possibility of producing the desired di-photon signal in LRSM 
with additional scalar multiplets, the neutral component of which can give rise to some fraction of total dark matter abundance in our Universe. 
It should be noted that around $26\%$ of present Universe's energy density is in the form of dark matter and its relic abundance observed 
by the Planck experiment \cite{Planck13} can be quoted as
\begin{equation}
\Omega_{\text{DM}} h^2 = 0.1187 \pm 0.0017
\label{dm_relic}
\end{equation}
The annihilation cross-section of thermal dark matter particles required to generate such relic abundance is of the order of weak interaction cross-sections leading to the so called \textit{WIMP Miracle}. In the above equation \eqref{dm_relic}, $\Omega$ is the density parameter and $h = \text{(Hubble Parameter)}/100$ 
is a parameter of order unity. In order to have a good production cross-section of the 750 GeV neutral scalar, we consider a specific version of LRSM \cite{mohaextLR} 
where one can have a 750 GeV neutral scalar with sizeable couplings to standard model quarks. We show that for the minimal particle content of this model, the reported 
di-photon signal can not be produced. We then incorporate septuplet pairs into the model whose neutral components can be a dark matter candidates. 
Due to the existence of triply, doubly and singly charged scalars, these septuplet pairs can significantly enhance the partial decay width of the 750 GeV 
scalar into di-photons. We show that for two pairs of such septuplets, the di-photon signal can be produced for maximal perturbative couplings and 
masses of charged septuplet components close to (but greater than)  375 GeV. For such large couplings, the relic abundance of septuplet dark matter 
remains subdominant requiring additional particle content which can give rise to dominant component of dark matter. Though subdominant, such dark matter 
candidates can however have promising signatures at dark matter indirect detection experiments. We also show that, for such choices of parameters and 
masses, the 750 GeV scalar has a total decay width of around 36 GeV, which will soon go through scrutiny at LHC experiment.

This paper is organized as follows. In section \ref{sec1}, we briefly discuss the version of LRSM we adopt in this work and calculate the possibility 
of a 750 GeV neutral scalar in view of the LHC signal. In section \ref{sec2}, we extend this LRSM with pairs of septuplets and discuss the implications 
for di-photon signal. In section \ref{sec3}, we demonstrate how the neutral component of scalar septuplet can be sub-dominant component of dark 
matter. In section \ref{sec:landau}, we comment on the high scale validity of the model and then finally conclude our results in Sec\,\ref{sec4}.

\section{Left-Right Symmetric Model}
\label{sec1}
Left-Right Symmetric Model \cite{lrsm,lrsmpot} is one of the very well motivated BSM frameworks where the gauge symmetry of the electroweak theory is extended to $SU(3)_c \times SU(2)_L \times SU(2)_R \times U(1)_{B-L}$. The model not only explains the origin of parity violation in low energy weak interactions naturally, but also gives rise to light neutrino masses naturally through seesaw mechanisms. Parity violation at low energy occurs due to spontaneous breaking of left-right symmetry at high scale and seesaw origin of light neutrino mass is guaranteed due to natural inclusion of heavy right handed neutrinos as parts of $SU(2)_R$ fermion doublets. Here the difference between the baryon and lepton number is a local gauge symmetry and the electric charge formula 
relating baryons and leptons as follows
\begin{align}
Q=T_{3L}+T_{3R}+\frac{B-L}{2} 
\end{align}
\begin{table}[htb]
\begin{center}
\begin{tabular}{|c|c|c|c|c|}
\hline
Particle & $ SU(3) $ & $SU (2)_L$ & $SU(2)_R$ & $U(1)_{B-L}$ \\ 
\hline 
$q_{L}=\begin{pmatrix}u_{L}\\
d_{L}\end{pmatrix}$ & 3 & 2 & 1 & $\frac{1}{3}$\\ 
$q_{R}=\begin{pmatrix}u_{R}\\
d_{R}\end{pmatrix}$ & 3 & 1 & 2 & $\frac{1}{3}$\\ 
$\ell_{L}=\begin{pmatrix}\nu_{L}\\
e_{L}\end{pmatrix}$ & 1 & 2 & 1 & -1 \\ 
$\ell_{R}=\begin{pmatrix}\nu_{R}\\
e_{R}\end{pmatrix}$ & 1 & 1 & 2 & -1 \\ 
\hline
$\Phi=
\left(\begin{array}{cc}
\ \phi^0_{11} & \phi^+_{11} \\
\ \phi^-_{12} & \phi^0_{12}
\end{array}\right)
$ & 1 & 2 & 2 &0 \\ 
$\Delta_L =
\left(\begin{array}{cc}
\ \delta^+_L/\surd 2 & \delta^{++}_L \\
\ \delta^0_L & -\delta^+_L/\surd 2
\end{array}\right)$ & 1 & 3 & 1 & 2 \\
$\Delta_R =
\left(\begin{array}{cc}
\ \delta^+_R/\surd 2 & \delta^{++}_R \\
\ \delta^0_R & -\delta^+_R/\surd 2
\end{array}\right)$ & 1 & 1 & 3 & 2\\
\hline 
\end{tabular}
\end{center}
\caption{Particle content of the minimal LRSM.}
\label{table1}
\end{table}

The particle content and their transformation under the gauge symmetry of LRSM $SU(3)_c \times SU(2)_L\times SU(2)_R \times U(1)_{B-L}$ are shown in table \ref{table1}. In the symmetry breaking
pattern, the neutral component of the Higgs triplet $\Delta_R$ acquires a vev to break the gauge symmetry of the LRSM into that of the SM and then to the $U(1)$ of electromagnetism by the vev of the neutral component of Higgs bidoublet $\Phi$:
$$ SU(2)_L \times SU(2)_R \times U(1)_{B-L} \quad \underrightarrow{\langle
\Delta_R \rangle} \quad SU(2)_L\times U(1)_Y \quad \underrightarrow{\langle \Phi \rangle} \quad U(1)_{em}$$
After the symmetry breaking, four neutral scalars emerge, two from the bidoublet $(H^0_0, H^0_1)$, one from right handed triplet $(H^0_2)$ and another from left handed triplet $(H^0_3)$. Similarly there are two neutral pseudoscalars, one from the bidoublet $(A^0_1)$ and another from the left handed triplet $(A^0_2)$. Among the charged scalars, there are two singly charged ones $(H^{\pm}_1, H^{\pm}_2)$ and two doubly charged ones $(H^{\pm \pm}_1, H^{\pm \pm}_2)$. Here, $H^0_0$ can be identified as SM like Higgs of mass 125 GeV. In order to avoid the flavor changing neutral currents (FCNC) processes, the neutral scalars from  bi-doublet $H^0_1, A^0_1$ have to be heavier than 10 TeV \cite{fcnc}. This leaves only the neutral scalars from triplets $(H^0_2, H^0_3)$ to be a candidate for 750 GeV neutral resonance. As shown in \cite{Dasgupta:2015pbr}, this neutral scalar originating from the triplets can not give rise to the desired signal.

The reduced di-photon cross-section through 750 GeV $H^0_{2,3}$ in MLRSM is partly due to the small partial decay width $\Gamma (H^0_{2,3} \rightarrow \gamma \gamma)$ as well as small production cross-section $\sigma ( pp \rightarrow H^0_{2,3})$ of $H^0_{2,3}$ in proton proton collisions at LHC. Since $H^0_{2,3}$ do not directly couple to quarks, it can be produced only through its mixing with the standard model Higgs, suppressing the cross-section by the mixing angle squared. If the extra neutral scalar of 750 GeV mass were from the bidoublet, the production cross-section can be enhanced due to its direct coupling with quarks. Fortunately, some interesting extension of MLRSM is possible which can allow a neutral scalar with 750 GeV mass originating from the bidoublet \cite{mohaextLR}. Some other alternatives to evade FCNC in these models have been proposed in \cite{altLR}. The model proposed by \cite{mohaextLR} includes the new particles shown in table \ref{table2}.
\begin{table}[htb]
\begin{center}
\begin{tabular}{|c|c|c|c|c|}
\hline
Particle & $ SU(3) $ & $SU (2)_L$ & $SU(2)_R$ & $U(1)_{B-L}$ \\ 
\hline 
$\rho=
\left(\begin{array}{cc}
\ \rho^+_{1} & \rho^{++} \\
\ \rho^0 & \rho^+_{2}
\end{array}\right) $ & 1 & 2 & 2 & 2 \\
$ Q^{\prime} = \left(\begin{array}{c}
T \\
t^{\prime}
\end{array}\right)$ & 3 & 1 & 2 & $\frac{7}{3}$ \\
\hline 
\end{tabular}
\end{center}
\caption{Additional particle content of the model \cite{mohaextLR} in addition to the particles in minimal LRSM.}
\label{table2}
\end{table}

Here $\rho$ is an extra bidoublet scalar and $Q^{\prime}$ is a vector like $SU(2)_R$ quark doublet with exotic $U(1)_{B-L}$ charge $7/3$. Since the model was assumed to break discrete left-right symmetry (or D parity) at very high scale by a parity odd singlet scalar, the vector like $SU(2)_L$ quarks were considered to be very heavy and hence decoupled from the rest of the particles. Additional $Z_4$ symmetry was assumed in order to forbid simultaneous couplings of the standard model quarks to $\Phi$ and $\tilde{\Phi} = \sigma_2 \Phi^* \sigma_2$. 

The essential feature of the left-right symmetric theories is to naturally provide the light neutrino masses via type-I 
plus type-II seesaw mechanisms as follows
\begin{equation}
	m_\nu = M_L - m_D  M^{-1}_R m^T_D
	= m_\nu^{II} + m_\nu^I\,,
\label{neutrino-mass}
\end{equation}
where $m_D$ is the Dirac neutrino mass connecting light LH and heavy RH neutrinos, $M_L$ ($M_R$) is the Majorana neutrino 
mass for LH (RH) neutrinos. When $SU(2)_R$ gauge symmetry breaking and discrete left-right symmetry breaking occur at different scales (possible in a D-parity broken class of left-right models) the type II seesaw contribution 
is negligible leaving only dominant type-I seesaw contributions to light neutrino masses.

\subsection{750 GeV neutral scalar in LRSM}
As discussed in \cite{mohaextLR}, this extended LRSM has a richer Higgs sector compared to the MLRSM. The three neutral components of two bidoublets $\Phi, \rho$ acquire vev  $k_1, k_2, v_{\rho}$ which are responsible for electroweak symmetry breaking and hence related by
$$ (k^2_1+k^2_2+v^2_{\rho})^{1/2} =v^2= 246 \; \text{GeV}$$
The neutral component of $\Delta_R$ also acquires a vev $v_R$. Due to mixing of scalars through quartic couplings, the neutral scalar mass matrix is in general non-diagonal. Using the expressions for mass matrix elements given in \cite{mohaextLR} we diagonalise the mass matrix in order to arrive at the desired mass spectra. The values of the vev are chosen as 
$$ k_1 = 183 \; \text{GeV}, k_2 = 36.6 \; \text{GeV}, v_{\rho} = 160 \; \text{GeV}, \; v_R = 3305 \; \text{GeV}$$
where the ratio $k_1/k_2=5$ is chosen in order to be in agreement with flavor data, shown in details by \cite{mohaextLR}. Choosing suitable values of scalar potential parameters, we can arrive at the mixing matrix
\begin{equation}
U=
\left(\begin{array}{cccc}
0.7647 & 0.0801 & 0.6378 & -0.0432 \\
0.618 & -0.356 & -0.70 & -0.007 \\
0.170 & 0.930 & -0.323 & -0.010 \\
0.039 & 0.011 & 0.019 & 0.998 \\
\end{array}\right)
\nonumber 
\end{equation}
in the basis $(k_1, k_2, v_{\rho}, v_R)$. The same choices of parameters also give the lighter scalar masses 
as $m_h=125$ GeV and $m_H=750$ GeV respectively. Thus the physical neutral scalars will be $\text{Re}(\phi^0_1, 
\phi^0_2, \rho^0, \delta^0_R)$ upto a rotation by the mixing matrix $U$ given above. There are two neutral pseudo 
scalars originating from two bidoublets in the model, one of which can be as light as 750 GeV as shown by \cite{mohaextLR}. 
If we consider the left handed triplet scalar $\Delta_L$, then there is one more neutral scalar and one more neutral 
pseudo scalar  in the spectrum. Apart from the neutral scalars and pseudo scalars, there are three singly charged 
scalars and two doubly charged scalars around TeV scale. Including $\Delta_L$ at low energy increases the number 
of singly and doubly charged scalars by one.

Due to its coupling with the standard model quarks, the 750 GeV neutral scalar $H$ can be produced with a cross-section similar to to the standard model like Higgs $h$ with 750 GeV mass. The dominant production channel is the 
gluon gluon fusion with top quarks in a loop. The production cross-section $\sigma (pp\rightarrow H)$ is same as 
$\sigma (pp \rightarrow h)$ upto a factor decided by the ratio of effective coupling of top quarks to $h, H$ 
respectively. The 750 GeV scalar $H$ after production, can decay at tree level into standard model fermions 
(dominantly to top quark pairs), electroweak vector bosons W, Z and standard model Higgs $h$. Due to many 
tree level decay channels, this Higgs $H$ can possibly have a large decay width, as suggested by the 
di-photon resonance search at LHC.
\begin{figure}[!h]
\centering
\begin{tabular}{cc}
\epsfig{file=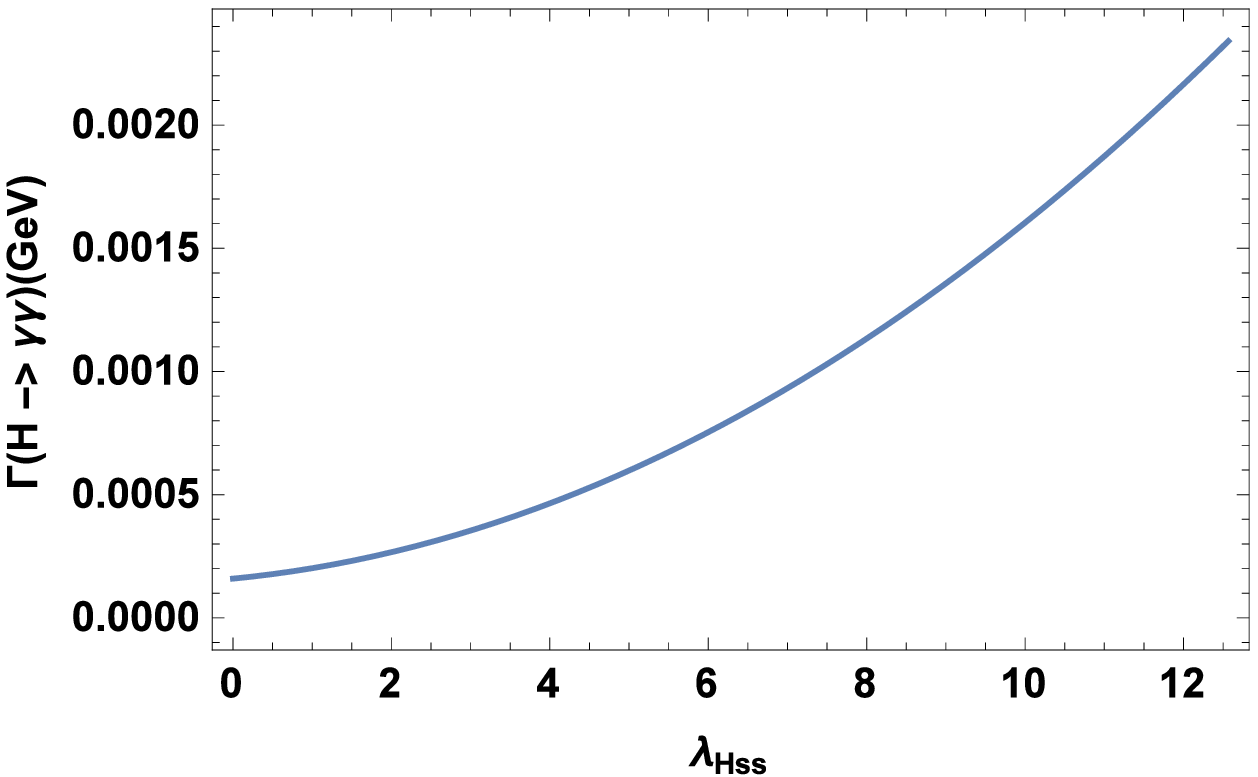,width=0.5\textwidth,clip=} &
\epsfig{file=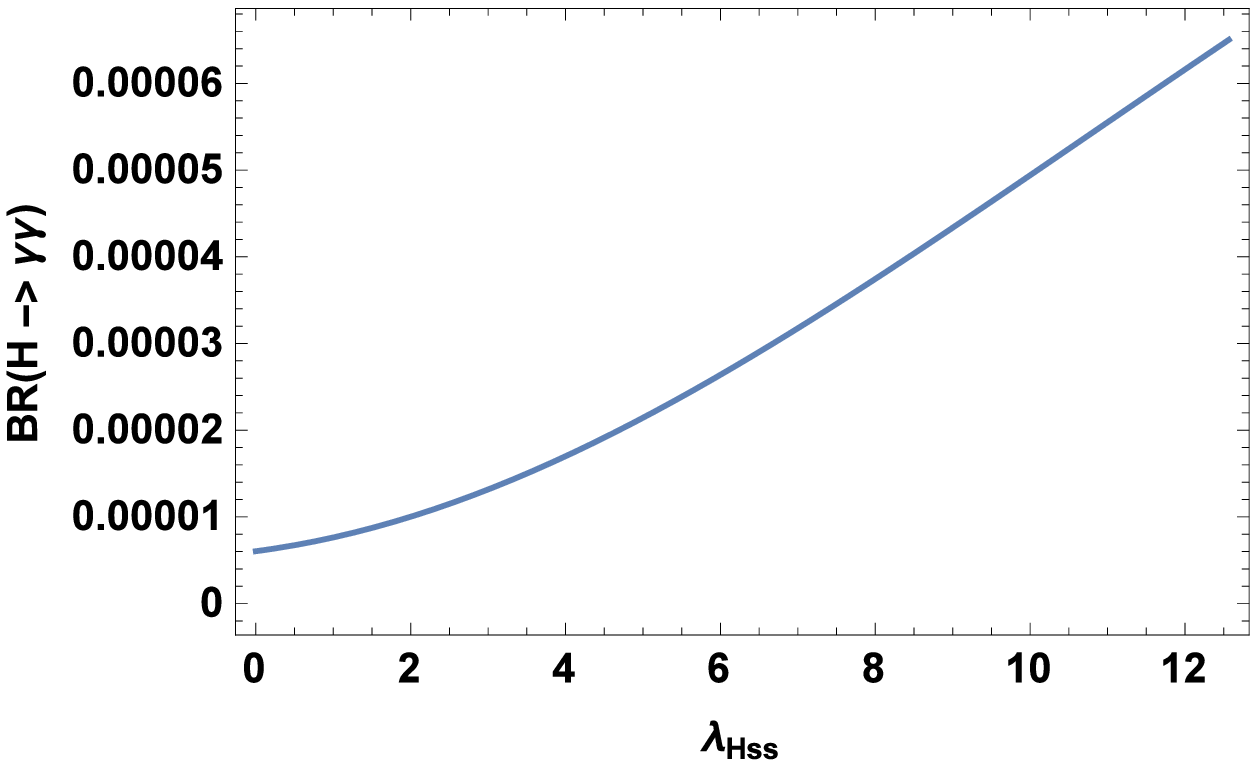,width=0.5\textwidth,clip=} 
\end{tabular}
\caption{Partial decay width and branching ratio into di-photons in LRSM}
\label{fig1}
\end{figure}
\begin{figure}[!h]
\centering
\epsfig{file=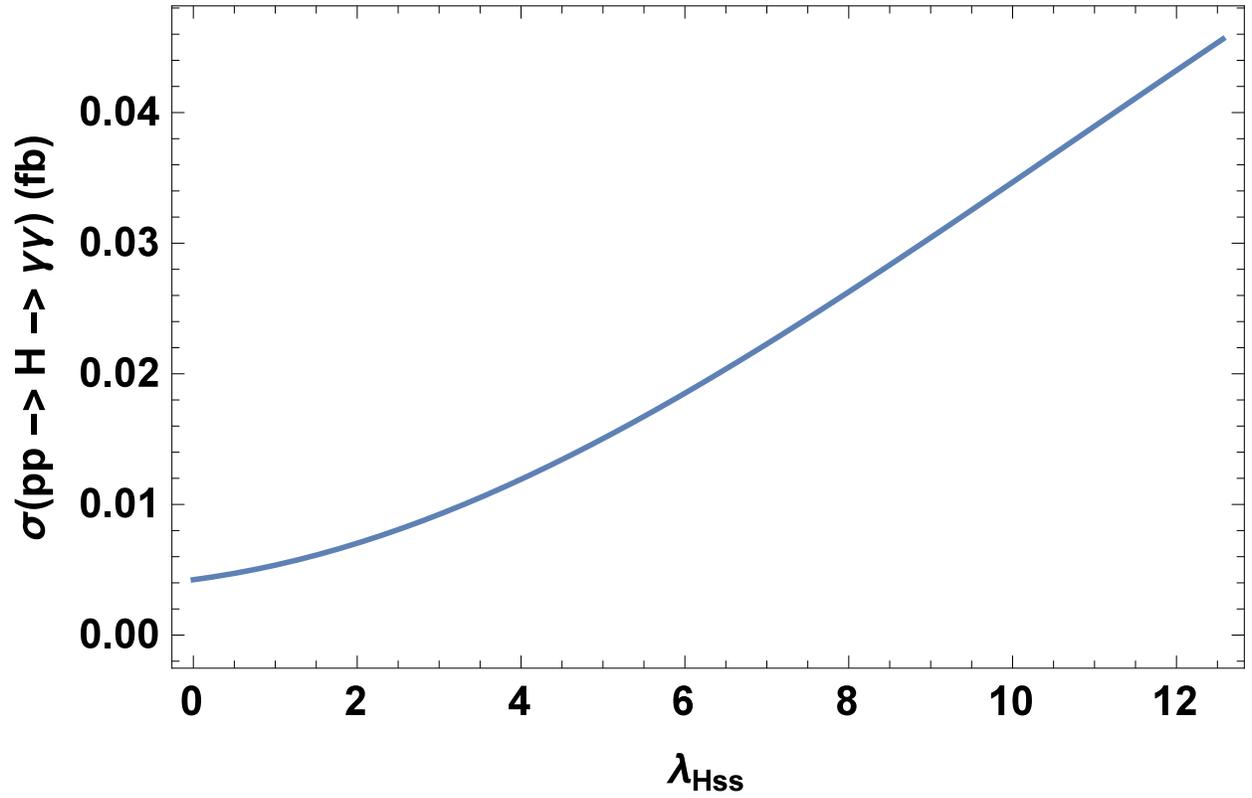,width=1.0\textwidth,clip=}
\caption{Production cross-section of di-photons in LRSM}
\label{fig2}
\end{figure}

The model briefly discussed above has the following charged particles which can go in a loop allowing 
the 750 GeV scalar $H$ decaying into two photons:
\begin{itemize}
\item Standard model charged fermions: all the quarks and charged leptons.
\item Electroweak charged bosons $W^{\pm}$
\item Singly charged and doubly charged scalars originating from scalar bidoublets and triplets.
\end{itemize}
The corresponding partial decay width of $H$ to di-photons can be calculated using the formula \cite{physrepdjouadi, di-photonwidth}
\begin{equation}
\Gamma (H\rightarrow \gamma \gamma) = \frac{G_F \alpha^2 m^3_H}{128 \sqrt{2} \pi^3} \lvert \sum_f N_c Q^2_f 
\frac{g_{Hff}}{g_{hff}} A^H_{1/2} (\tau_f) +\frac{g_{HWW}}{g_{hWW}} A^H_1 (\tau_W) + Q^2_{s} g_{Hss} A^H_0 (\tau_s) \rvert^2
\end{equation}
where $G_F$ is Fermi coupling constant, $\alpha$ is fine structure constant, $N_c$ is the color factor of 
charged fermion in loop, $Q_{f,s}$ are electromagnetic charges of fermions and scalars in loop and $\tau_i  
= m^2_H/4m^2_i$ with $i$ running over all charged particles in loop. The form factors $A^H_{1/2, 1, 0}$ 
for fermion, vector boson and scalar respectively are given by 
$$ A^H_{1/2} (\tau) = 2[ \tau + (\tau-1)f(\tau) ] \tau^{-2} $$
$$ A^H_1 (\tau) = -[2\tau^2 + 3\tau + 3(2\tau-1) f(\tau) ] \tau^{-2} $$
$$ A^H_0 (\tau) = -[ \tau -f(\tau) ] \tau^{-2} $$
with the function $f(\tau)$ is given by
\[ f(\tau) = 
\begin{cases}
\text{arcsin}^2 \sqrt{\tau}, & \tau \leq 1 \\
-\frac{1}{4} \left ( \log \frac{1+\sqrt{1-\tau^{-1}}}{1-\sqrt{1-\tau^{-1}}} -i\pi \right )^2, & \tau >1
\end{cases}
\]
Among the charged fermions, the top quark loop will dominate the partial decay width of $H$ to $\gamma \gamma$. 
The effective coupling of $H$ to top quarks in the model is $g_{Hff} = \sqrt{2} (V_L M^{\prime \prime}_u V^{\dagger})_{33}/v$
where $V_{L,R}$ are left and right handed quark mixing matrices and $M^{\prime \prime}_u$ is the 
effective up-type quark mass matrix originating from $H$ contribution to the masses. For the above 
choice of values of vev, $(V_L M^{\prime \prime}_u V^{\dagger})_{33}$ comes out to be approximately 
$-159$ GeV, which we use in our calculations. The coupling of $H$ to $W$ boson pairs is given by 
$$g_{HWW} =\frac{g^2}{2} (U_{21} k_1+U_{22} k_2 +U_{23} v_{\rho})$$
where $U_{24}, v_R$ are not appearing as $\Delta_R$ does not couple to $SU(2)_L$ gauge bosons. 
The coupling of $H$ to charged scalar is 
$$ g_{Hss} = -\frac{m_W}{g m^2_s} \lambda_{Hss} (U_{21} k_1+U_{22} k_2 +U_{23} v_{\rho}+U_{24}v_R) $$
assuming all the charged scalars to have a same effective dimensionless coupling $ \lambda_{Hss} $ to $H$. 
In the expression for decay width, $g_{hff}, g_{hWW}$ are the corresponding light standard model Higgs $h$ 
couplings to fermions and gauge bosons respectively.

It can be seen from figure \ref{fig1}, that for largest possible perturbative value of $\lambda_{Hss} = 4\pi$, 
one can achieve a partial decay width $\Gamma (H \rightarrow \gamma \gamma) \approx 0.0024$ GeV and corresponding 
branching ratio $\text{BR} (H \rightarrow \gamma \gamma) \approx 8 \times 10^{-5}$. We also make an estimate 
of the production cross-section $\sigma ( pp \rightarrow H \rightarrow \gamma \gamma)$ and show its variation 
with $\lambda_{Hss}$. We take the production cross-section of standard model like Higgs with mass 750 GeV 
to be 850 fb \cite{SigmaH} and multiply it by the factor $(V_L M^{\prime \prime}_u V^{\dagger})_{33}/m_t$ 
squared to take care of the difference in $htt$ and $Htt$ couplings. The resulting di-photon production 
cross-section is shown in figure \ref{fig2}. Thus, for largest possible values of $\lambda_{Hss}$, the maximum 
cross-section of the di-photon signal is approximately 0.045 fb, around two order of magnitudes below 
the reported signal of LHC. 

For the same choices of couplings, we also check the contributions of these charged scalars to the standard 
model like Higgs $h$ decaying into di-photons. We find that $\Gamma^{\text{new}}(h \rightarrow \gamma \gamma) 
\approx 4 \times 10^{-6}$ for maximal dimensionless couplings. This value is already around $25\%$ of standard 
model partial decay width $\Gamma^{\text{SM}}(h \rightarrow \gamma \gamma)$ and hence will be ruled out by 
8 TeV LHC data \cite{hgg8TeV}.
\begin{figure}[!h]
\centering
\begin{tabular}{cc}
\epsfig{file=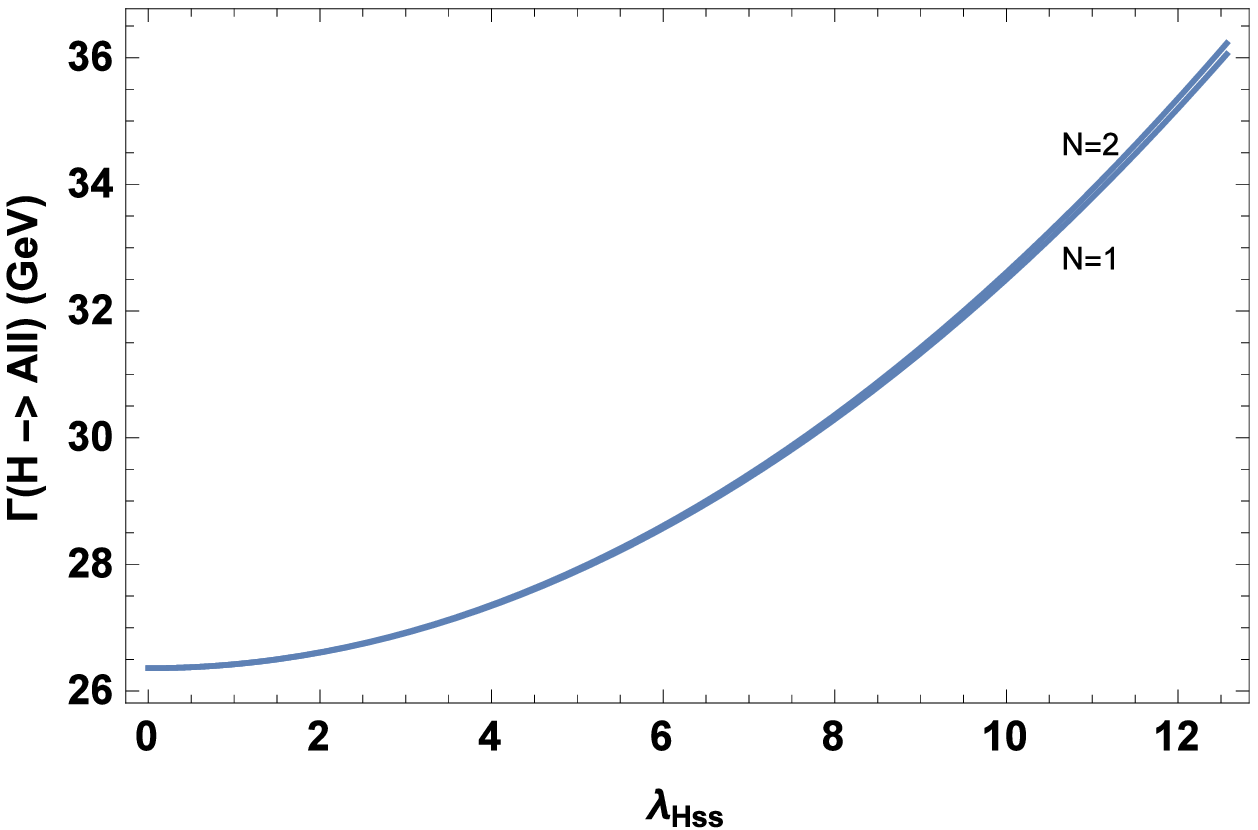,width=0.5\textwidth,clip=} &
\epsfig{file=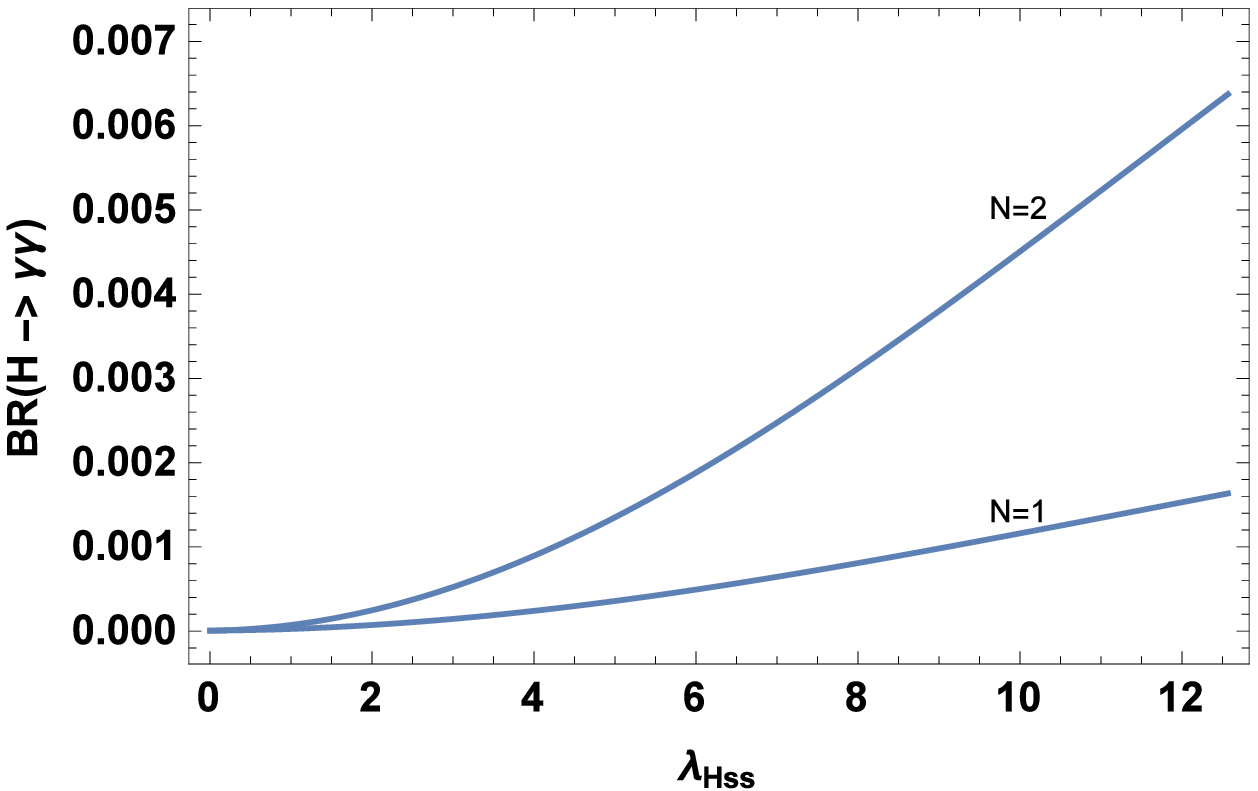,width=0.5\textwidth,clip=} 
\end{tabular}
\caption{Total decay width of $H$ and di-photon production cross-section in LRSM with N septuplet pairs}
\label{fig3}
\end{figure}
\begin{figure}[!h]
\centering
\epsfig{file=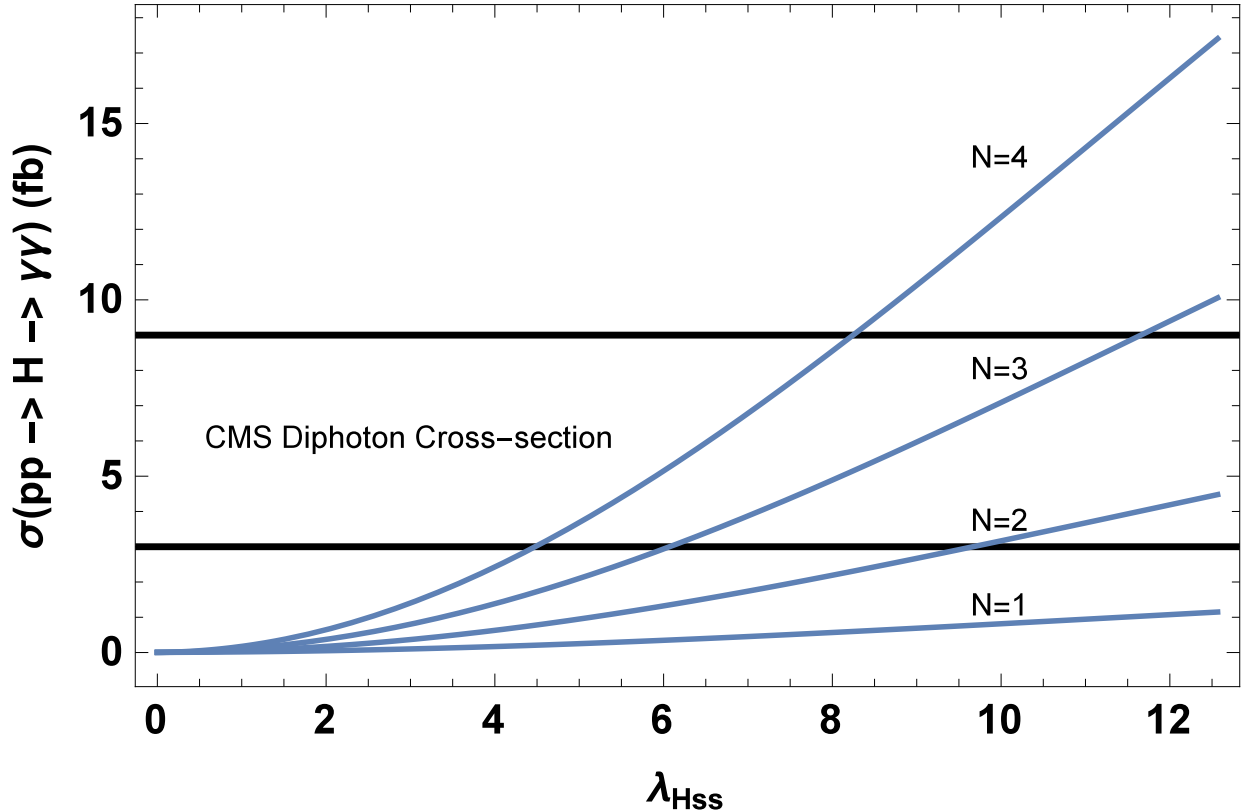,width=1.0\textwidth,clip=}
\caption{Production cross-section of di-photons in LRSM with N septuplet pairs}
\label{fig31}
\end{figure}
\section{Di-photon Excess in LRSM with Septuplets}
\label{sec2}
\begin{table}[htb]
\begin{center}
\begin{tabular}{|c|c|c|c|c|}
\hline
Particle & $ SU(3) $ & $SU (2)_L$ & $SU(2)_R$ & $U(1)_{B-L}$ \\ 
\hline 
$S_L $ & 1 & 7 & 1 & 0 \\
$ S_R$ & 1 & 1 & 7 & 0 \\
\hline 
\end{tabular}
\end{center}
\caption{Real septuplets added to LRSM.}
\label{table3}
\end{table}

To enhance the partial decay width of 750 GeV scalar $H$ into di-photons without conflicting with 
$\Gamma^{\text{SM}}(h \rightarrow \gamma \gamma)$ we include a pair of septuplets into the above model. The transformation of these septuplets are shown in table \ref{table3}. The motivation for inclusion of  septuplet scalars is two fold: 1) its charged 
components enhance the $H\to \gamma \gamma$ branching 
fraction for di-photon excess signal, 
2) its neutral component can be stable dark matter candidate. However, we shall restrict ourselves to those 
parameters space of the scalar 
septuplet which is helpful in explaining 
the di-photon signal even if it contributes marginally to the relic abundance of dark matter. The introduction 
of additional
dark matter components can compensate 
so as to give the correct number needed for relic density of dark matter of the Universe. The usual structure 
of a  scalar septuplet is given by
\begin{align}
S_{L,R} = (S_{L,R}^{+++},\,S_{L,R}^{++},\, S_{L,R}^{+},\, S_{L,R}^{0},\,- S_{L,R}^{-},\,S_{L,R}^{--},\,-S_{L,R}^{---})^T
\end{align}

Since the  septuplet has only $SU(2)_{L,R}$~dimensions, it can interact with the corresponding gauge 
bosons in following way
\begin{align}
\mathcal{L}^{\rm int}_{S_L, S_R} &= i g_{L,R} W^3_{L,R,\mu} (S^{-}_{L,R} \overset{\longleftrightarrow}{\del^\mu} S_{L,R}^{+} +2 \,S^{--}_{L,R} \overset{\longleftrightarrow}{\del^\mu} S_{L,R}^{++} + 3\, S^{---}_{L,R} \overset{\longleftrightarrow}{\del^\mu} S_{L,R}^{+++} ) \nonumber \\
&\quad+ \left[ i g_{L,R} W^-_{L,R,\mu} (\sqrt{6}\, S^{0}_{L,R} \overset{\longleftrightarrow}{\del^\mu} S_{L,R}^{+} +\sqrt{5} \,S^{-}_{L,R} \overset{\longleftrightarrow}{\del^\mu} S_{L,R}^{++} + \sqrt{3}\, \phi^{--}_{L,R} \overset{\longleftrightarrow}{\del^\mu} S_{L,R}^{+++} ) +\hc \right] \nonumber \\
&\quad+ g_{L,R}^2 W^3_{L,R,\mu}W^{3,\mu}_{L,R}\left( | S_{L,R}^+|^2 + 4\, | S_{L,R}^{++}|^2 + 9\, | S_{L,R}^{+++}|^2\right) \nonumber \\
&\quad+ g_{L,R}^2 W^+_{L,R,\mu}W^{-,\mu}_{L,R}\left(6\, (S_{L,R}^0)^2 +11\, | S_{L,R}^+|^2 + 8\, | S_{L,R}^{++}|^2 + 3\, | S_{L,R}^{+++}|^2\right) \nonumber \\
&\quad+ g_{L,R}^2 \left[ W^-_{L,R,\mu}W^{-,\mu}_{L,R}\left(\sqrt{30}\, S_{L,R}^0 S_{L,R}^{++} + \sqrt{15}\, S_{L,R}^- S_{L,R}^{+++}-3\, S_{L,R}^+ S_{L,R}^{+}\right) +\hc \right] \nonumber \\
&\quad+ g_{L,R}^2 \left[ W^3_{L,R,\mu}W^{-,\mu}_{L,R}\left(\sqrt{6}\, S_{L,R}^0 S_{L,R}^{+} + \sqrt{45}\, S_{L,R}^- S_{L,R}^{++}+\sqrt{75}\, S_{L,R}^{--} S_{L,R}^{+++}\right) +\hc \right] .
\end{align}
where we defined $A\overset{\longleftrightarrow}{\del^\mu}B \equiv A\del^\mu B - B\del^\mu A$. Also the scalar sector 
extended with the following scalar interactions
\begin{align}
\mathcal{L}^{\rm scalar}(S_L, S_R)  &= \sum_{ L,R} \left[\frac12 (D_\mu S_{L,R})^\dagger D^\mu S_{L,R} -\frac12 M^2 | S_{L,R}|^2 -\sum_{k=1,2} \lambda_{k} [S_{L,R}]^4_k \right] -\lambda_{LR} 	| S_L|^2 | S_R|^2 \nonumber \\
&\quad- \sum_{ L,R} \left[\lambda_\Phi |\Phi|^2 |S_{L,R}|^2 + \lambda_\rho |\rho|^2 |S_{L,R}|^2 \right] \nonumber \\
&-\sum_{ L,R} \left[\lambda_{\Delta_2} |\Delta_{L,R}|^2 | S_{L,R}|^2+\lambda_{\Delta_3} (\Delta_{L,R}^\dagger 
\Delta_{L,R})  (S_{L,R}^\dagger S_{L,R}) \right] ,
\label{eq:septuplet_lagrangian}
\end{align}
The interesting point to note here is that the mass-splitting between charged and neutral components arise through electroweak radiative corrections and is found to be $M^Q_{S_L}-M^0_{S_L} \simeq \unit[170] {MeV} Q^2$, for the left handed septuplet \cite{strassler}. Thus, it is clear that the 
lightest component is a neutral one. For the right handed multiplet, the mass splitting can be written as \cite{Heeck:2015qra}
\begin{equation}
M_Q-M_0 \approx \frac{\alpha_2 M}{4\pi} Q^2 [ f(r_{W_R})-c^2_M f(r_{Z_2})-s^2_W s^2_M f(r_{Z_1})-s^2_W f(r_{\gamma}) ]
\end{equation}
where $s_M = \sin{\theta_M} = \tan{\theta_W}$ \cite{thetawm}, $r_X = M_X/M$ and 
$$ f(r) = 2 \int^1_0 dx (1+x) \log{[x^2+(1-x)r^2]}$$
Thus, the splitting between charged and neutral components of right handed septuplets 
depend upon the right handed gauge boson masses. Larger splitting can be introduced by incorporating additional scalar fields. Such splitting has very important relevance in computation of dark matter relic abundance.

Due to the existence of triply, doubly and singly charges scalars in a septuplet multiplet, the decay of $H$ into 
di-photons can be substantially enhanced. Assuming the septuplet pairs $S_{L,R}$ to have no coupling with the bidoublet 
$\rho$, we can significantly keep the light Higgs $h$ coupling to the septuplet scalars under control. The septuplet 
scalar couplings to the light Higgs $h$ is given by 
$$ g_{hss} = -\frac{m_W}{g m^2_s} \lambda_{hss} (U_{11} k_1+U_{12} k_2 +U_{14} v_R) $$
which is almost zero for the above choices of vev's and mixing matrix. However, the coupling of 750 GeV scalar to the 
septuplet scalars can still be sizeable
$$ g_{Hss} = -\frac{m_W}{g m^2_s} \lambda_{Hss} (U_{21} k_1+U_{22} k_2 +U_{24}v_R) $$
which is approximately equal to $-\frac{m_W}{g m^2_s} (77)$. Thus, choosing the mixing angles of the mixing matrix $U$ 
appropriately, one can keep the new physics contributions to $\Gamma (h\rightarrow \gamma \gamma)$ under control while 
enhancing the $\Gamma (H \rightarrow \gamma \gamma)$ simultaneously. The effective dimensionless couplings $\lambda_{hss}, 
\lambda_{Hss}$ are approximately same, as both $h$ and $H$ are from the same bidoublet and we are assuming identical 
septuplet couplings to bidoublet and triplet.

We now consider the new physics contribution to $\Gamma (H \rightarrow \gamma \gamma)$ to be coming only from the 
charged components of the septuplets. The contributions from charged scalars in bidoublets and triplet are assumed 
to be negligible, as enhancing their contributions also increases the $\Gamma (h\rightarrow \gamma \gamma)$ as seen 
above. Taking the septuplet and $t, W$ contributions to the partial decay width of $H$ to di-photons, we plot the 
total decay width of $H$ and branching ratio $\text{BR}(H \rightarrow \gamma \gamma)$ in figure \ref{fig3}. We also 
plot the production cross-section $\sigma ( pp \rightarrow H \rightarrow \gamma \gamma)$ in figure \ref{fig31}. 
We have considered two different cases: one with a pair of septuplets and another with two pairs of septuplets. 
Although the total decay width remains more or less same in both cases, the di-photon production cross-section 
can be enhanced in the latter, to be in agreement with LHC data. Here we have considered the total decay width to 
be
$$ \Gamma^{\text{Total}}_H = \Gamma^{t\bar{t}}_H + \Gamma^{WW}_H +\Gamma^{ZZ}_H +\Gamma^{hh}_H + \Gamma^{\gamma \gamma}_H 
+ \Gamma^{gg}_H$$
where 
$$ \Gamma^{t\bar{t}}_H = \frac{3}{16\pi} \lvert y_{Ht\bar{t}} \rvert^2 m_H \left ( 1-\frac{4m^2_t}{m^2_H} \right)^{3/2} \approx 25.87 \; \text{GeV}$$
$$\Gamma^{WW}_H = \frac{1}{8\pi} \frac{g^2_{HWW}}{m_H} \big [ 1+\frac{1}{2}\left (1-\frac{m^2_H}{2m^2_W} \right )^2 \big] \left(1-\frac{4m^2_W}{m^2_H} \right )^{1/2} \approx 0.29 \; \text{GeV} $$
$$\Gamma^{ZZ}_H = \frac{1}{16\pi} \frac{g^2_{HZZ}}{m_H} \big [ 1+\frac{1}{2}\left (1-\frac{m^2_H}{2m^2_Z} \right )^2 \big] \left(1-\frac{4m^2_Z}{m^2_H} \right )^{1/2} \approx 0.09 \; \text{GeV}$$ 
$$\Gamma^{gg}_H = \frac{G_F \alpha^2_s m^3_H}{36 \sqrt{2} \pi^3} \lvert y_{H t\bar{t}} \rvert^2 \lvert \frac{3}{4} A^H_{1/2} (\tau_t) \rvert^2 \approx 0.053 \; \text{GeV}$$
The partial decay widths $\Gamma^{hh}_H, \Gamma^{\gamma \gamma}_H $ vary with $hH$ 
and septuplet-H couplings. One interesting prediction of the model is the total decay width of around 36 GeV for maximal 
coupling $\lambda_{Hss}$, which the future LHC data should be able to confirm or rule out. 

\section{Dark Matter in LRSM with Septuplets}
\label{sec3}
The neutral component of the scalar septuplet required for enhancing $\Gamma (H\to \gamma \gamma)$ to explain the 
di-photon excess can be 
a stable dark matter candidate. The neutral component of the scalar septuplet is found to be accidentally stable 
in the minimal dark matter spirit by not allowing any renormalizable couplings which leads to the decay because 
of high $SU(2)$-dimension \cite{Cirelli:2005uq,Cirelli:2015bda,Garcia-Cely:2015dda}. It should however be noted that, such accidental stability may not be respected by unknown physics at a high energy scale, say the scale of quantum gravity. Such new physics can induce the decay of these dark matter candidates with a lifetime much less than the age of the Universe. It was shown in \cite{luzio}, by considering an effective field theory framework that higher dimensional operators can lead to fast decay of septuplet dark matter particle. One can guarantee the stability of septuplet dark matter only when some additional gauge symmetries are responsible for its exact stability. We do not pursue this topic further in this work and leave it for a future study.

Since the maximal di-photon excess can be achieved for charged septuplet masses close to 375 GeV, we consider low mass 
septuplet dark matter. For dark matter stability, the neutral components of septuplets should be lighter than the charged 
ones. Such a mass splitting can be achieved by considering either loop corrections or introducing some other multiplets 
whose non-zero vev can induce a large mass difference. Typically, the loop corrected splittings are small, which may 
allow co-annihilations between charged and neutral components of septuplets during the time of freeze-out. Here we 
consider large splitting so that the coannihilation effects can be neglected. We independently calculate the relic 
abundance of left and right handed septuplet dark matter. For left handed septuplet $S^0_L$, the dominant annihilation 
channel is $S^0_L S^0_L \rightarrow W^+ W^-$ for $M_{\text{DM}} \geq M_W$. The annihilation cross section is given by \cite{Cirelli:2005uq}
\begin{equation}
\langle \sigma v \rangle \approx \frac{1}{64\pi M^2_{\text{DM}} g_s} \{ g^2 (3-4n^2+n^4)+16Y^4g^4_Y+8g^2 g^2_Y Y^2 (n^2-1) \}
\end{equation}
where $n=7$ is the multiplet dimension, $g_s = 2n (n)$ for complex (real) multiplet and $Y$ is the hypercharge which is zero for a real septuplet. Apart from that there are also annihilation 
channels through the 750 GeV scalar H into the standard model particles. The annihilation cross sections we consider in this work are given by
\begin{equation}
\sigma (\text{DM} \; \text{DM} \rightarrow f \bar{f}) = \frac{N_c}{16\pi s} \lvert y_{Hf\bar{f}} \rvert^2 f^2_{Hss} \frac{s-4m^2_f}{(s-m^2_H)^2+\Gamma^2_H m^2_H} \left ( \frac{s-4m^2_f}{s-4M^2_{\text{DM}}} \right )^{1/2}
\end{equation}
\begin{equation}
\sigma (\text{DM} \; \text{DM} \rightarrow h h) = \frac{1}{32\pi s} \frac{f^2_{Hss} f^2_{Hhh}}{(s-m^2_H)^2+\Gamma^2_H m^2_H} \left ( \frac{s-4m^2_h}{s-4M^2_{\text{DM}}} \right )^{1/2}
\end{equation}
\begin{equation}
\sigma (\text{DM} \; \text{DM} \rightarrow V V) = \frac{1}{16\pi s g_V}  \frac{f^2_{Hss} f^2_{HVV}}{(s-m^2_H)^2+\Gamma^2_H m^2_H}  \big [ 1+\frac{1}{2}\left (1-\frac{m^2_H}{2m^2_V} \right )^2 \big]\left ( \frac{s-4m^2_V}{s-4M^2_{\text{DM}}} \right )^{1/2}
\end{equation}
where $f_{Hss} = \lambda_{\text{DM}} (U_{21} k_1+U_{22} k_2 +U_{24}v_R)$ is the dark matter coupling to 750 GeV neutral scalar $H$, $f_{Hhh} = \lambda_{Hhh} (U_{21} k_1+U_{22} k_2 +U_{23} v_{\rho}+U_{24}v_R)$ is the $hhH$ coupling, $V$ denotes electroweak massive bosons $W, Z$ with $g_W = 1, g_Z = 2$ and $f_{HVV}$ is same as $g_{HWW}$ defined earlier. These annihilation cross sections can be used to calculate the relic abundance of dark matter given by \cite{Jungman:1995df}
\begin{equation}
\Omega_{\chi} h^2 \approx \frac{3 \times 10^{-27} \text{cm}^3 \text{s}^{-1}}{\langle \sigma v \rangle}
\label{eq:relic}
\end{equation}

\begin{figure}[!h]
\centering
\begin{tabular}{cc}
\epsfig{file=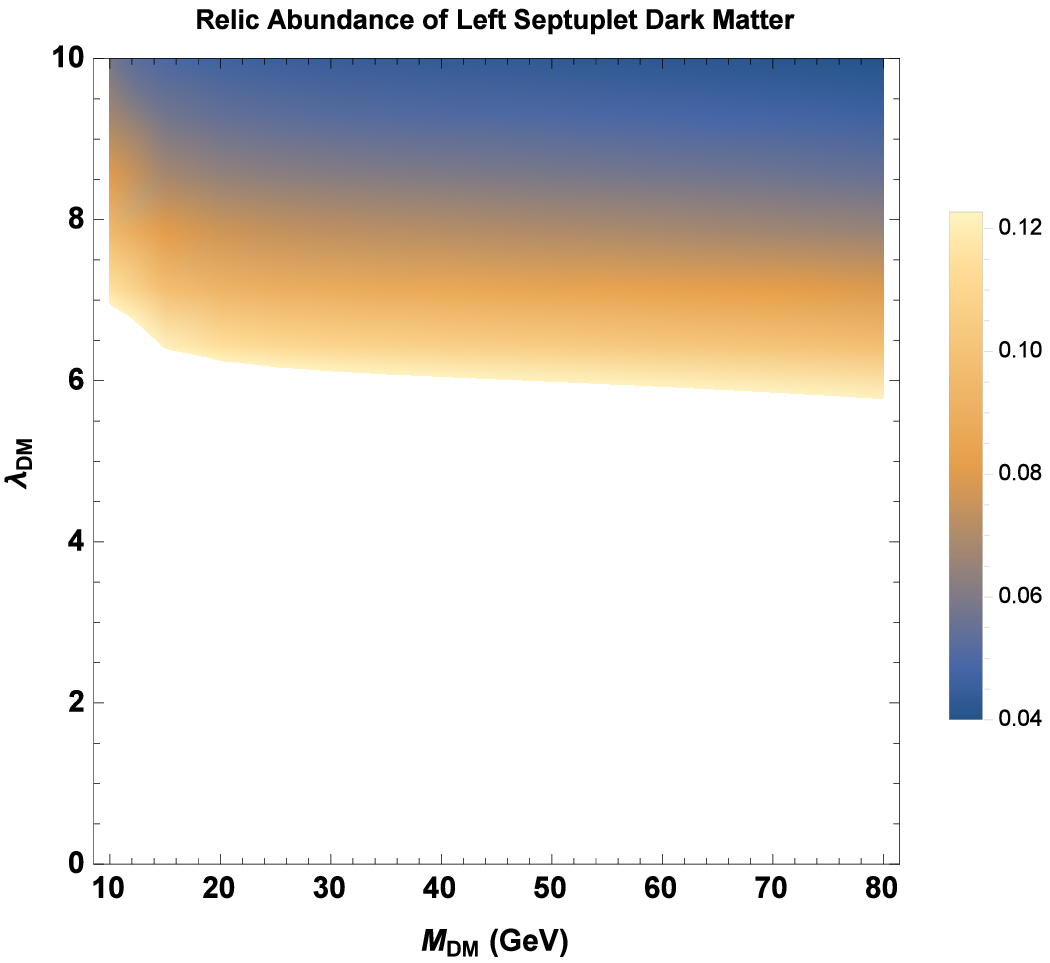,width=0.5\textwidth,clip=} &
\epsfig{file=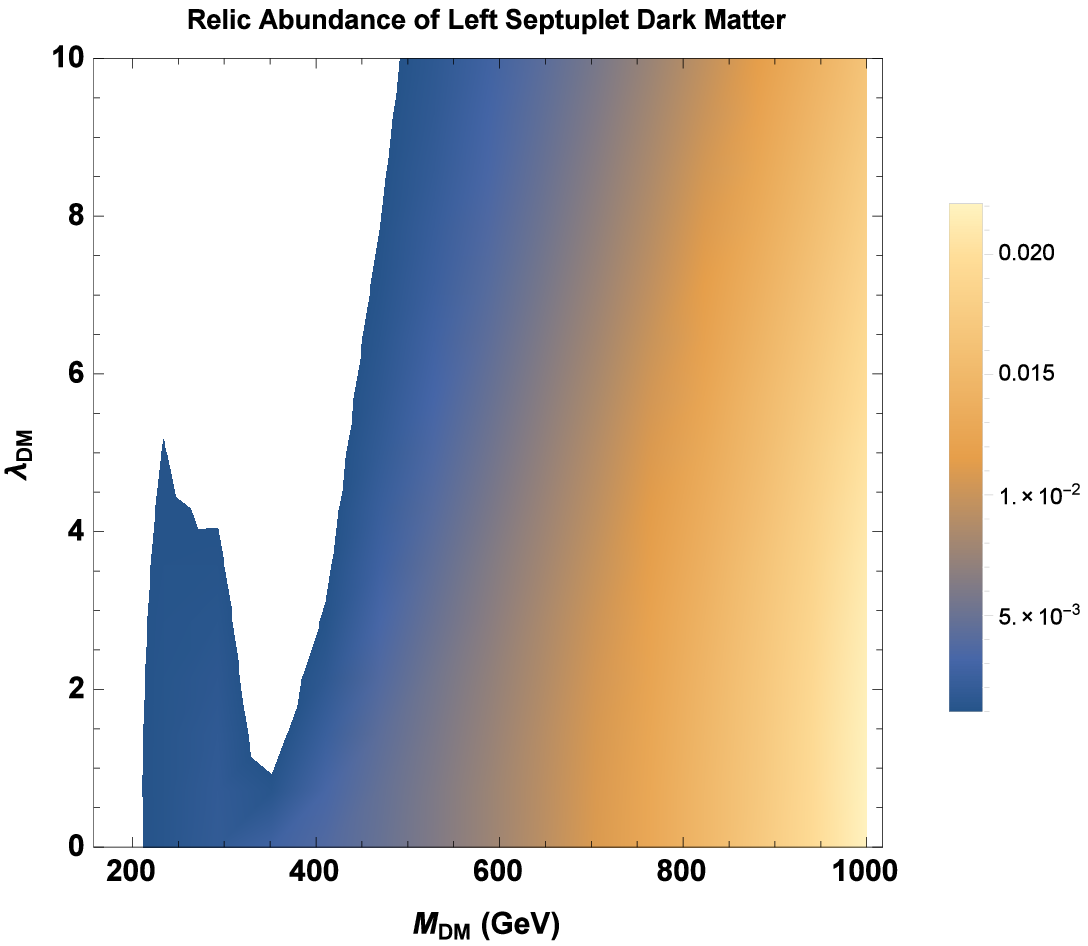,width=0.5\textwidth,clip=} 
\end{tabular}
\caption{Parameter space giving rise to subdominant relic abundance of left handed septuplet dark matter}
\label{fig4}
\end{figure}

\begin{figure}[!h]
\centering
\epsfig{file=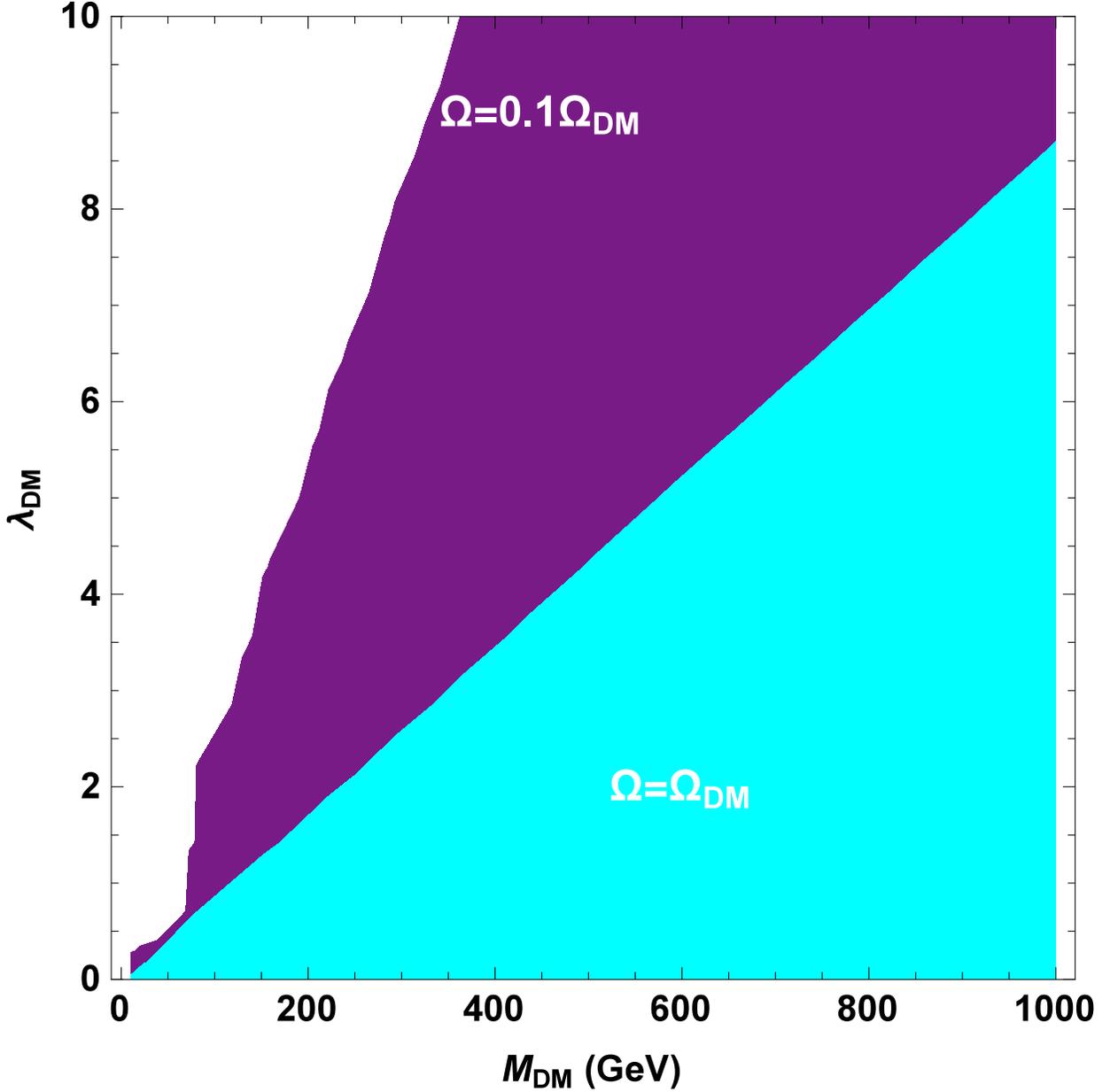,width=1.0\textwidth,clip=} 
\caption{Parameter space giving rise to maximum value of H mediated direct detection cross-section allowed by LUX experiment for dominant and subdominant septuplet dark matter}
\label{fig5}
\end{figure}

Annihilations through light Higgs $h$ are negligible due to tiny coupling of septuplets with $h$. It is observed that, 
for $M_{\text{DM}} \geq M_W$, the relic abundance of $S^0_L$ is suppressed whereas for $M_{\text{DM}} < M_W$, the relic 
density can be sufficient. The relevant parameter space for both the mass range is shown in figure \ref{fig4}. The low 
mass regime, although produces correct relic abundance, will be ruled out by the direct detection constraints as seen 
from figure \ref{fig5}. The dark matter nucleon scattering can be mediated by the same 750 GeV neutral scalar which appear in the self-annihilations. Therefore, the same DM-H coupling $\lambda_{\text{DM}}$ that appears in annihilation cross sections also arise in spin-independent scattering cross section. The most stringent constraints on dark matter nucleon spin independent scattering cross-section 
comes from the LUX experiment \cite{LUX}. We consider the minimum upper limit on the dark matter-nucleon spin independent 
cross-section from LUX experiment \cite{LUX} which is $7.6 \times 10^{-46} \; \text{cm}^2$ and show the allowed region of 
parameter space in figure \ref{fig5} for both dominant and subdominant dark matter.
\begin{figure}[!h]
\centering
\epsfig{file=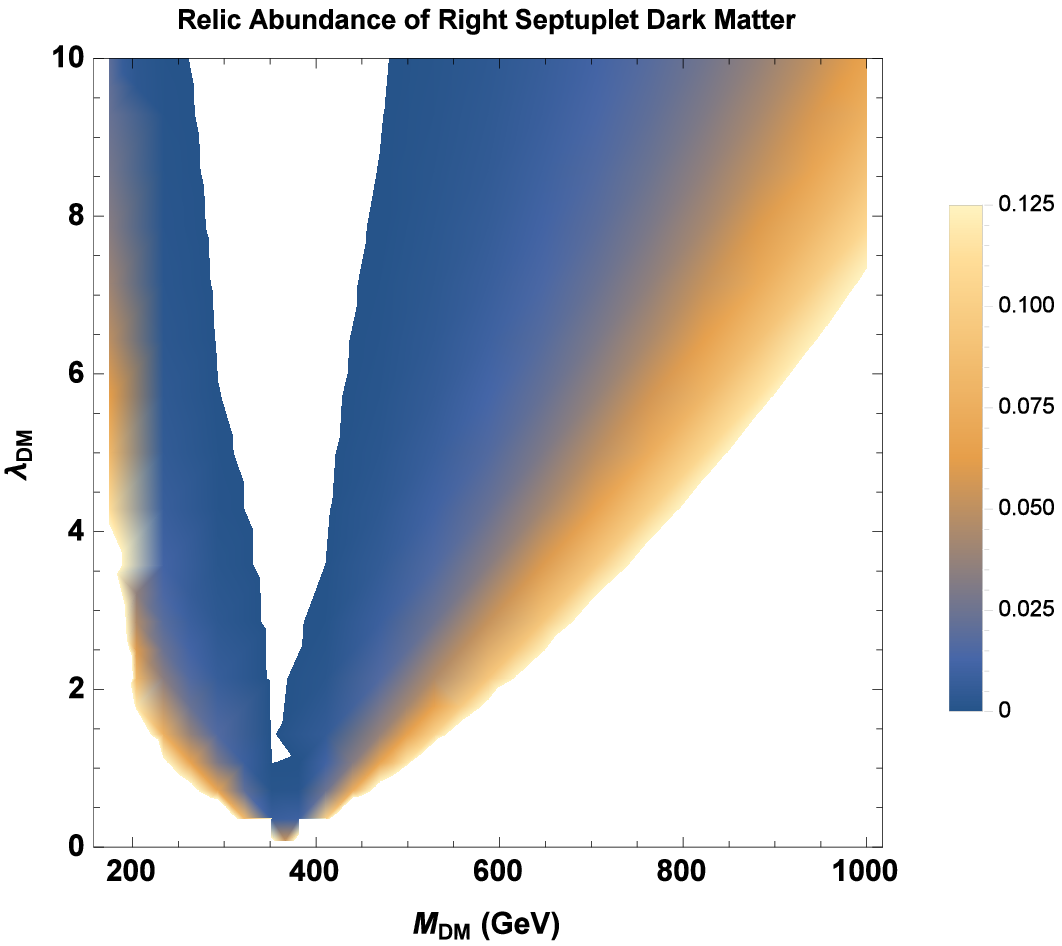,width=1.0\textwidth,clip=} 
\caption{Parameter space giving rise to relic abundance of right handed septuplet dark matter}
\label{fig6}
\end{figure}

Considering the right handed gauge bosons to be very heavy, the right handed septuplet dark matter relic abundance can be calculated by considering only the $H$ mediated channels into standard model particles. The relevant parameter space is shown in figure \ref{fig6}. It can be seen that for dark matter masses below 375 GeV, the coupling $\lambda_{\text{DM}}$ between H and $S^0_R$ needs to be low $\lambda_{\text{DM}} \leq 2$ in order to satisfy constraints from relic abundance and direct detection simultaneously. However, to get the required di-photon signal, the coupling of $H$ with septuplets are required to be high as shown above. As seen from figure \ref{fig5}. such high values of coupling will give rise to subdominant dark matter in this model. For such subdominant dark matter, the direct detection constraints will also be relaxed by the ratio of its abundance to actual dark matter abundance in the Universe. As seen from figure \ref{fig5}, for a dark matter component which gives rise to $10\%$ of total 
dark matter abundance, the constraints from direct detection becomes weaker, allowing more regions of parameter space. As discussed in \cite{Garcia-Cely:2015dda}, such septuplet dark matter in spite of being subdominant, can have promising signatures at indirect detection experiments searching for gamma rays from dark matter annihilations. Present experimental bounds in fact rule out left handed septuplet masses upto 20 TeV if it gives rise to total dark matter abundance. For lower masses of left handed septuplet, the experimental bounds constrains its relic abundance to be less than $1-10\%$ of total dark matter abundance.

Since the scalar septuplet contribution to the total dark matter of the Universe is subdominant, there should be alternative ways 
to explain the correct dark matter fraction consistent with the Planck data. We briefly discuss the possible new dark matter multiplets which can 
compensate for the remaining dominant dark matter fraction. We provide a few possibilities of dark matter candidates whose stability is either ensured automatically 
because of high $SU(2)$~dimension which forbids the tree level decay or exactly stable because of a remnant $$\mathcal{Z}_2\simeq (-1)^{B-L}$$ 
symmetry arising after the spontaneous symmetry breaking of LRSM down to SM gauge group i.e, $SU(2)_R \times U(1)_{B-L} \to U(1)_Y$. Under this 
remnant discrete symmetry $\mathcal{Z}_2\simeq (-1)^{B-L}$, the usual leptons are odd while all bosons including scalars and gauge bosons are even. 
For detailed discussion, one may refer to \cite{Heeck:2015qra,Garcia-Cely:2015quu}.

More specifically, one can extend this discussion of remnant discrete symmetry $\mathcal{Z}_2$ to quarks in the following manner.  
Defining $X=3(B-L)$--in order to make integer $B-L$ charges for quarks--usual quarks carry charge $X(q_{L,R})=1$, 
leptons with $X(\ell_{L,R})=-3$ while scalars have $X-~$charge as $X(\Phi)=0$, $X(\rho)=6$, $X(\Delta_R)=6$. Thus, 
the breaking of $U(1)_X$ by scalar triplet $\Delta_R$ with $6$-unit of charge leads to 
$$\mathcal{Z}_6 = \mathcal{Z}_3 \times \mathcal{Z}_2\quad, \mbox{where} \mathcal{Z}_2= (-1)^{3(B-L)}\, \quad \mbox{for leptons}$$
While $\mathcal{Z}_3$ is assigned as a remnant discrete symmetry for quarks transforming as $q_{L,R} \to e^{i \pi/3} q_{L,R}$. 
It is quite clear now that all the fermions are odd while scalar bosons $\Phi, \rho, \Delta_R$ transforming are even 
under the remnant discrete symmetry. With this formulation, if we introduce a fermionic multiplet having $B-L=0$, i.e, 
$\mathcal{Z}_2({\rm DM})=(-1)^{3\times 0}=+1$ which implies that it is even under discrete symmetry. Thus, the neutral component 
is exactly stable ensured by this discrete symmetry as other known fermions are odd under this discrete symmetry. 

Using the formalism of dark matter stability, the neutral lightest components of each fermionic multiplets like 
$(3,1,0,1)_F+(1,3,0,1)_F$, $(5,1,0,1)+(1,5,0,1)$, $(2,2,0,1)_F$ and $(3,3,0,1)_F$ and the lightest neutral component of scalar 
multiplet while $H_L(2,1,-1)+H_R(1,2,-1,1)$ can be a viable dark matter candidate. Here we have two component dark matter for fermionic 
triplets and quintuplets as the neutral component of each of left- and right-handed multiplet is a stable dark matter candidate 
and thus, the relic abundance is sum of two components i,e $\Omega_{\rm DM} h^2\simeq \Omega^{L}_{\rm DM} h^2+\Omega^{R}_{\rm DM} h^2$.

\section{Landau Pole and High Scale Validity of the Model}
\label{sec:landau}
Landau Pole is a point where either the value of gauge coupling constant $g_i$ becomes infinite 
or the inverse fine structure constant $\alpha^{-1}=4\pi/g^2_i$ for a quantum field theory is zero. 
The focus is to examine whether the addition of additional septuplet scalars to the left-right symmetric 
model under consideration leads to hitting the Landau Pole or not. Since the septuplet scalars transform non-trivially under $SU(2)_{L,R}$ one needs to check the renormalization group (RG) running of gauge couplings for $SU(2)_{L,R}$ only. 

The renormalization group evolution (RGE) equations for gauge couplings at loop level is given by \cite{Gross:1973ju,Jones:1981we}
\begin{eqnarray}
\frac{d\, \alpha^{-1}_{i}}{d\, t}=-\frac{\pmb{b_i}}{2 \pi}
\label{rge-coupl}
\end{eqnarray}
where we denote $t=\ln(\mu)$, $\alpha_{i}=g^2_{i}/(4 \pi)$ as fine structure constants and 
$\pmb{b_i}$ being the one-loop beta coefficients for a given $i^{\rm th}$ gauge group. The 
formula for one-loop beta coefficients is given by
for $\pmb{b_i}$ is
\begin{eqnarray}
	&&\pmb{b_i}= - \frac{11}{3} \mathcal{C}_{2}(G) 
				 + \frac{2}{3} \,\sum_{R_f} T(R_f) \prod_{j \neq i} d_j(R_f) 
				 + \frac{1}{3} \sum_{R_s} T(R_s) \prod_{j \neq i} d_j(R_s).
\label{oneloop_bi}
\end{eqnarray}
where $\mathcal{C}_2(G)$ is defined as the quadratic Casimir operator for gauge bosons in their adjoint 
representation. The other parameters $T(R_f)$ and $T(R_s)$ are traces of the irreducible representation $R_{f,s}$ 
for a given fermion and scalar representation. Similarly, $d(R_{f,s})$ is defined as the dimension 
of a given representation $R_{f,s}$. 

More specifically, one can write down the RG evolution equations for $SU(2)_L$ fine-structure coupling $\alpha_2 \equiv g_2^2/4\pi$ 
at one-loop level from a scale $\lambda$ to $\Lambda >\lambda$: \cite{Gross:1973ju,Jones:1981we}
\begin{align}
\frac{1}{\alpha_2 (\Lambda)} = \frac{1}{\alpha_2 (\lambda)} - \frac{b_2}{2\pi}\log \left(\frac{\Lambda}{\lambda}\right) \, .
\end{align}
Thus, the Landau pole -- $\alpha_2^{-1} (\Lambda_\text{LP}) =0$ -- is given by 
\begin{align}
\Lambda_\text{LP} \simeq \lambda \exp \left[ \frac{2\pi}{b_2} \alpha_2^{-1} (\lambda) \right] ,
\end{align}
if $b_2 >0$. Here $\lambda$ is the left-right symmetry breaking scale around few TeV range, 
$b_2$ is the one-loop beta coefficient for $SU(2)_{L,R}$ for the present model with septuplet 
scalars motivated to explain the recent diphoton excess around $750~$GeV reported by ATLAS and CMS.

The fermions and scalar content of the present left-right symmetric model with gauge group 
$SU(2)_L \times SU(2)_R \times U(1)_{B-L} \times SU(3)_C$ is given by
\begin{align}
\mbox{Fermions:}&\quad q_{L}(2,1,1/3,3)\, ,\quad  q_{R}(1,2,1/3,3)\, , \nonumber \\
    &\quad \ell_{L}(2,1,-1,1)\, ,\quad  \ell_{R}(1,2,-1,1)\, , \nonumber \\
\mbox{Scalars:}&\quad \Phi(2,2,0,1)\, ,\quad  \rho(2,2,2,1)\, , \nonumber \\
    &\quad \Delta_{L}(3,1,2,1)\, ,\quad  \Delta_{R}(1,3,2,1)\, , \nonumber
\end{align}
plus additional vector-like fermion $Q^\prime(1,2,7/3,3)$ whose RG effects is neglected here. 
In order to know the value of $SU(2)_{L,R}$ gauge couplings, one examine the RGEs from 
SM to LRSM and derived the couplings at left-right breaking scale (say at 5 TeV). The RG running results 
$\alpha^{-1}_{2L}\simeq 31.5$ and $\alpha^{-1}_{2R}\simeq 48.5$ at $5~$TeV. With the inclusion of scalar septuplets 
$(7,1,0,1) + (1,7,0,1)$ at left-right breaking scale and using the derived one-loop beta coefficients  $b_{2L}=20/3(2), b_{2R}=22/3(8/3)$ \cite{Hamada:2015bra} where the numbers in the parentheses are for real scalar septuplets, 
the Landau poles are found to be 
\begin{align}
&\Lambda_\text{LP} \simeq 5 \times 10^{16}\, \mbox{GeV}\, \quad \mbox{for} \; SU(2)_L \nonumber \\
&\Lambda_\text{LP} \simeq 5 \times 10^{22}\, \mbox{GeV}\, \quad \mbox{for} \; SU(2)_R\, \nonumber
\end{align}
For four multiplets, the Landau pole occurs at lower energy $\Lambda_{\text{LP}} = 10^9 \; \text{GeV} (SU(2)_L), \Lambda_{\text{LP}} = 10^{14} \; \text{GeV} (SU(2)_R)$. Thus, the introduction of higher $SU(2)_{L,R}$ multiplets like septuplet scalars in the present model can 
severely modifies the running of inverse of fine structure constants $\alpha^{-1}_{2L,2R}$ and even can results 
in a Landau Pole at a scale below the scale of gravity or Planck scale. However, the safe condition that $\Lambda_\text{LP} > M_{\rm Pl}\simeq 1.2 \times 10^{19}$ 
GeV can be achieved by incorporating the presence of new dynamics at high energy scale which we do not pursue here.
\begin{figure}[!h]
\centering
\epsfig{file=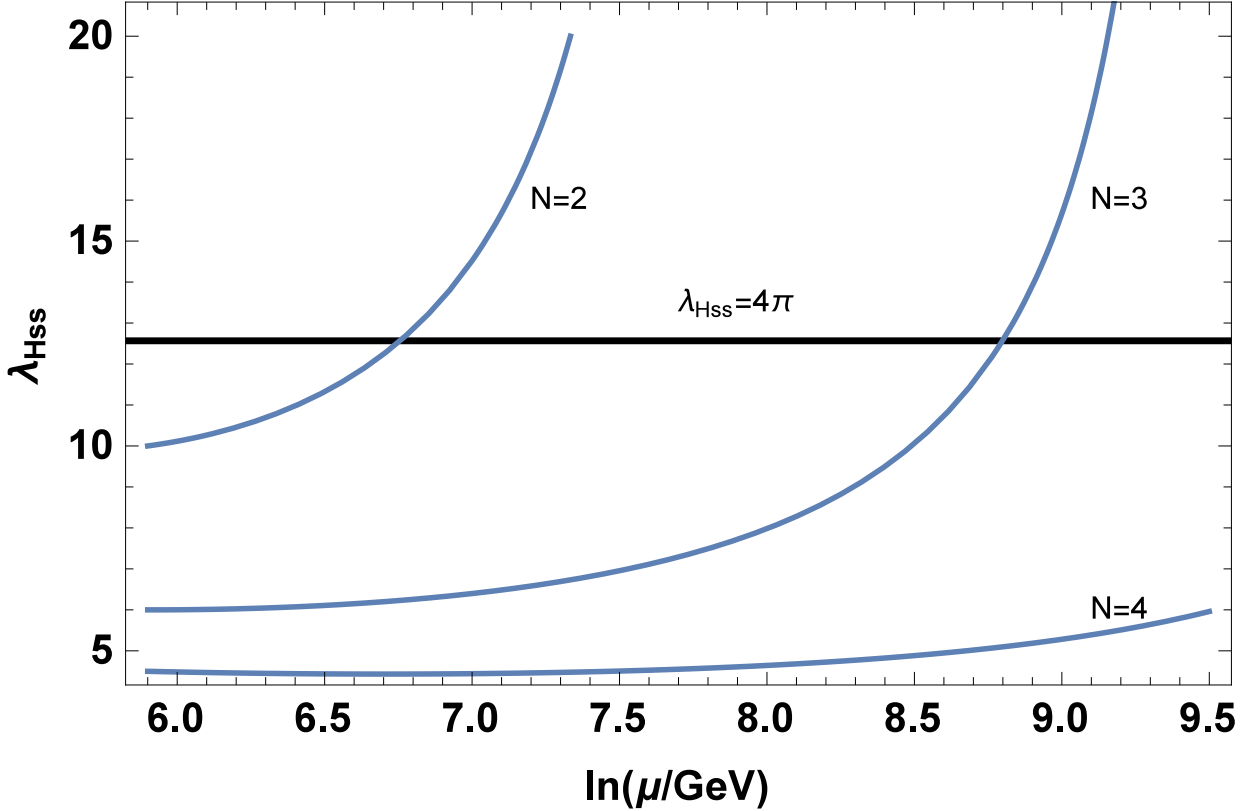,width=1.0\textwidth,clip=}
\caption{RG evolution of septuplet-750 GeV Higgs coupling}
\label{figRGE}
\end{figure}
While studying the effects of septuplets on the RG running of gauge couplings, it is equally important to check the RG evolution of scalar potential couplings. Among them, the coupling between the septuplet and the 750 GeV scalar is very important, as one needs to choose a large initial value of it in order to satisfy the LHC data on di-photon cross section, as seen from figure \ref{fig31}. The RGE equations for the scalar potential couplings in an extension of SM with real septuplets are given in \cite{Hamada:2015bra}. Assuming our low energy model below the scale of left-right symmetry as SM plus septuplets and the 750 GeV Higgs, we consider the evolution of scalar potential coupling between septuplet and the 750 GeV Higgs denoted by $\lambda_{Hss}$ in the above discussions. As seen from figure \ref{fig31}, a very high value of coupling $\lambda_{Hss} \geq 10$ needs to be chosen in a model with $N=2$ pairs of septuplets in order to satisfy CMS data. This makes the theory non-perturbative at a scale around 1 TeV as seen from figure \ref{figRGE}. If we have $N=3$ pairs of septuplets, then the lower bound on scalar coupling is $\lambda_{Hss} \geq 6$ for which the theory remains perturbative upto around 8 TeV. This limit can even be increased for $N=4$ pairs of septuplets as evident from figure \ref{figRGE}. Although, more pairs of septuplets contribute more severely to the running of $SU(2)$ gauge coupling, we check that it remains perturbative well above the scale we find for perturbative nature of scalar potential couplings. In this simplified analysis, we however assume the initial values of other septuplet couplings to be zero. Above the scale of left-right symmetry, one needs to include the RGE equations for other gauge couplings as well include the contribution of them in the RGEs of scalar potential couplings. However, the order of estimate analysis made for SM with septuplets and 750 GeV Higgs will not change significantly by a more complete RGE analysis for LRSM which we have skipped here. Thus, with the inclusion of 3 pairs of septuplets, the theory can remain perturbative upto the scale of 8 TeV, inviting additional new dynamics beyond LRSM well within the reach of LHC center of mass energy 14 TeV. This is a generic feature of other LRSM explanation of 750 GeV di-photon excess with additional vector like quarks \cite{Dev:2015vjd} where a cut-off scale of around 10 TeV was obtained.

\section{Compatibility with 8 TeV data}
\label{sec:8TeV}
\begin{figure}[!h]
\centering
\epsfig{file=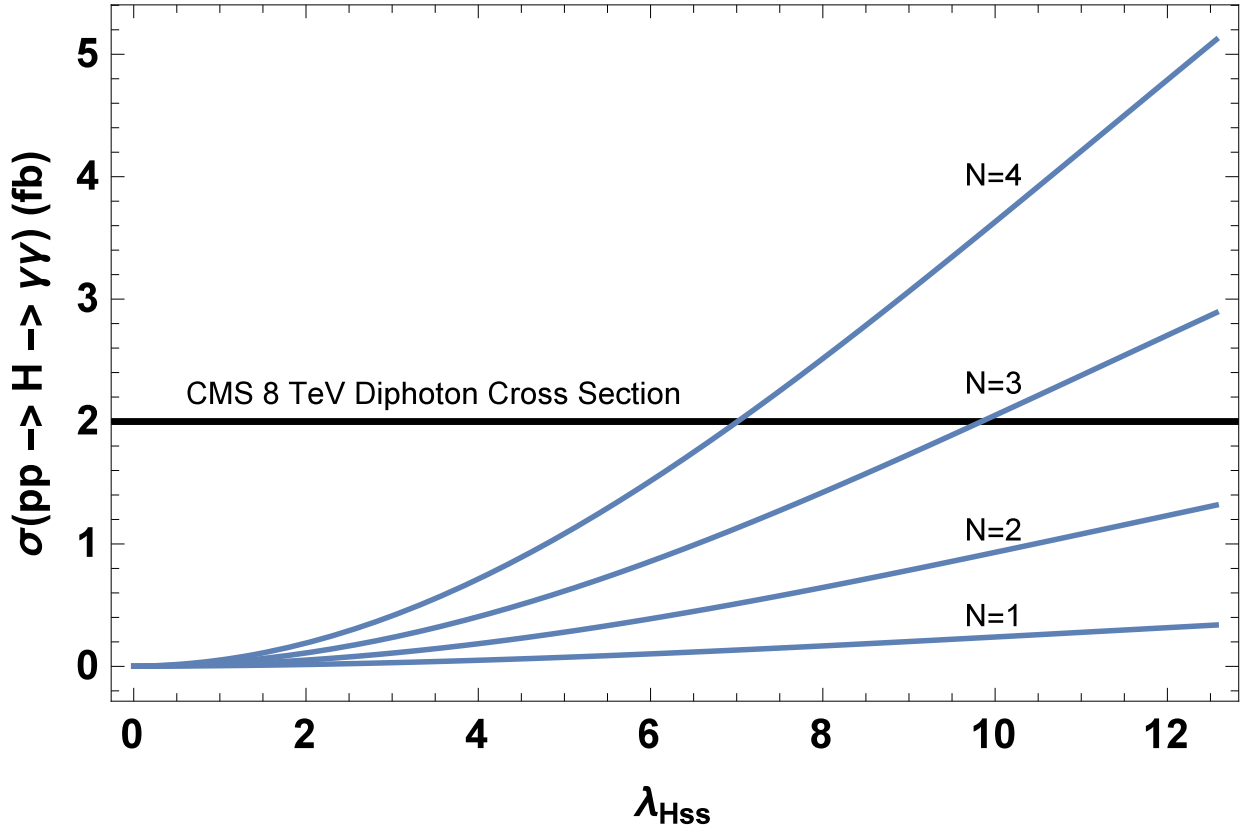,width=1.0\textwidth,clip=}
\caption{Production cross-section of di-photons in LRSM with N septuplet pairs, compared with 8 TeV LHC limits.}
\label{fig8tev}
\end{figure}

Considering the production cross section of SM like Higgs at 8 TeV LHC to be around 250 fb, we do the same analysis as before to get different final state production cross sections at 8 TeV. The results for diphoton final states at 8 TeV is shown in figure \ref{fig8tev}. It can be seen that for $N=3, 4$ pairs of septuplets, the 8 TeV data can constrain the dimensionless couplings to some range of values. The same range of values can however, give rise to the desired diphoton cross section at 13 TeV, as can be seen by comparing figure \ref{fig8tev} with \ref{fig31}. The 8 TeV CMS diphoton limit is taken from \cite{8TeVdiphoton}.

We also check the other relevant final states on which 8 TeV LHC data put strict bounds. For example, the cross section to WW final state is coming out to be 1.70 fb for maximum decay width of the 750 GeV scalar, around 36 GeV in our model. This lies way below the present limit of 47 fb \cite{WWCMS}.

One needs to collect more data in future LHC run before confirming the diphoton signal excess. Since 
the scalar resonance has other decay channels it is important to note down the other experimental constraints 
including the most relevant constraints from monojet plus missing transverse energy (MET) at LHC \cite{dijet-atlas,dijet-cms}. Many recent works explaining the 750 GeV diphoton excess have also addressed these bounds including \cite{Falkowski:2015swt}. In our model, the 750 GeV scalar can give rise to MET plus monojet through its couplings to septuplet dark matter. If dark matter mass $M_{DM} < 750/2$ GeV, the LHC constraints will severely constrain the couplings of dark matter with the 750 GeV scalar. This will show more preference towards heavier dark matter masses in our model.
\section{ Conclusion}
\label{sec4}
We have considered a specific class of left-right symmetric models where the additional neutral components of the bidoublet (apart from the SM Higgs) 
can be as light as 750 GeV and hence can be a potential candidate for the recently reported di-photon signal at LHC. Although the production cross-section 
of such a scalar can be comparable to the production of a SM like Higgs with 750 GeV mass, its branching ratio into di-photons remains too low to explain the LHC signal. 
In order to enhance the di-photon signal, we incorporate additional septuplet scalars whose charged components play the role in enhancing di-photon signal whereas the neutral components can play the role of subdominant dark matter candidates. We showed that for two pairs of such septuplets, the desired di-photon signal can be produced if certain dimensionless couplings are kept at maximal perturbative limit. The model also predicts the total decay width of the 750 GeV scalar to be around 36 GeV for such large values of couplings. This will definitely go through further scrutiny at future LHC measurements and have the potential of being falsified.

Interestingly, the same couplings which decide the di-photon production cross-section also play a role in deciding the relic abundance of the neutral component of these septuplets. We find that large values of such couplings suppress the dark matter relic abundance. The subdominant nature of the dark matter candidates is not only a requirement from the demand of producing the correct di-photon signal, but also a necessity from severe constraints coming from dark matter direct and indirect detection experiments. Being part of a very large $SU(2)$ multiplets, these dark matter candidates (in spite of being subdominant) can give rise to a significant indirect detection signals which can be probed by ongoing experiments. We have demonstrated that the inclusion of additional dark matter component can compensate this without affecting the di-photon phenomenology. We also check the high scale validity of the model by calculating the scale at which Landau pole occurs for the $SU(2)_{L,R}$ gauge couplings. We found that for four septuplet multiplets, the $SU(2)_{L}, SU(2)_R$ Landau pole occurs at $10^9, 10^{14}$ GeV respectively. This indicates the presence of additional new physics between the TeV scale LR symmetry and the scale of gravity or the Planck scale. Considering the RGEs of scalar potential couplings make even more strict constraint on the scale of validity of the theory demanding new dynamics beyond LRSM around 8-9 TeV. This leaves serious model building challenges for this type of scenarios, if the 750 GeV scalar hint survives further measurements and get established as the first beyond the SM discovery at LHC.

\section{Acknowledgements}
{\color{blue} DB and SS would like to thank P. Poulose, IIT Guwahati for very useful discussions. The work of SP is partly supported by DST, India under the financial grant 
SB/S2/HEP-011/2013.}


\end{document}